\documentclass[journal]{IEEEtran}
\IEEEoverridecommandlockouts
\usepackage{booktabs}
\usepackage{siunitx}
\usepackage{cite}
\usepackage{amsmath,amssymb,amsfonts}
\usepackage{graphics, framed}
\usepackage{graphicx, enumerate}
\usepackage{epsfig}
\usepackage[utf8]{inputenc}
\usepackage{epstopdf}
\usepackage{url}
\usepackage{multimedia}
\usepackage{paralist}
\usepackage[printonlyused]{acronym}

\usepackage{algorithmic}
\usepackage{textcomp}
\usepackage{float}
\usepackage{tabularx}
\usepackage{balance}
\usepackage{subfigure}
\usepackage{xcolor}
\def\BibTeX{{\rm B\kern-.05em{\sc i\kern-.025em b}\kern-.08em
T\kern-.1667em\lower.7ex\hbox{E}\kern-.125emX}}
\usepackage{mathtools}
\usepackage{cuted} 
\usepackage[export]{adjustbox}

\usepackage{dblfloatfix}
\usepackage{lipsum}
\usepackage{color,soul}
\usepackage{multirow}
\usepackage[overload]{textcase}
\newcommand{\FGR}[1]{Fig.~\ref{#1}}

\newcommand{\SEC}[1]{Section~\ref{#1}}

\acrodef{5G}[5G]{5\textsuperscript{th}-Generation}
\acrodef{BW}[BW]{bandwidth}
\acrodef{BER}[BER]{bit error rate}
\acrodef{IP}[IP]{Internet protocol}
\acrodef{BPSK}[BPSK]{binary phase-shift keying}
\acrodef{CW}[CW]{continuous wave}
\acrodef{CSI}[CSI]{channel state information}
\acrodef{D2D}[D2D]{device-to-device}
\acrodef{dB}[dB]{decibel}
\acrodef{dBi}[dBi]{decibel isotropic}
\acrodef{dBm}[dBm]{decibel over a milliwatt}
\acrodef{DSN}[DSN]{deep-space networks}
\acrodef{DTN}[DTN]{delay tolerant network}
\acrodef{Gbps}[Gbps]{gigabit per second}
\acrodef{GHz}[GHz]{gigahertz}
\acrodef{GAT}[GAT]{graph attention network}
\acrodef{THz}[THz]{Terahertz}
\acrodef{ISL}[ISL]{inter-satellite link}
\acrodef{RIS}[RIS]{Reconfigurable intelligent surfaces}
\acrodef{GM}[GM]{Gamma mixture}
\acrodef{PSK}[PSK]{phase shift keying}
\acrodef{QAM}[QAM]{quadrature amplitude modulation}
\acrodef{AWGN}[AWGN]{additive white Gaussian noise}
\acrodef{SNR}[SNR]{signal-to-noise ratio}
\acrodef{AF}[AF]{amplitude-and-forward}
\acrodef{MIMO}[MIMO]{multiple-input multiple-output}
\acrodef{mMIMO}[mMIMO]{massive-multiple-input multiple-output}
\acrodef{SDN}[SDN]{Software-defined network}
\acrodef{SON}[SON]{self-organizing network}
\acrodef{HetNet}[HetNet]{heterogeneous network}
\acrodef{FSO}[FSO]{free-space optics}
\acrodef{UM-MIMO}[UM-MIMO]{ultra-massive-MIMO}
\acrodef{AP}[AP]{access point}
\acrodef{UE}[UE]{user equipment}
\acrodef{NTN}[NTN]{non-terrestrial networks}
\acrodef{UAV}[UAV]{unmanned aerial vehicle}
\acrodef{HAPS}[HAPS]{high-altitude platform station}
\acrodef{LEO}[LEO]{low-Earth orbit}
\acrodef{BAN}[BAN]{body area network}
\acrodef{WLAN}[WLAN]{wireless local area network}
\acrodef{QoS}[QoS]{quality of service}
\acrodef{TCS}[TCS]{thermal control system}
\acrodef{QCL}[QCL]{quantum cascade laser}
\acrodef{CMOS}[CMOS]{complementary metal-oxide semiconductor}
\acrodef{V-HetNet}[V-HetNet]{vertical heterogeneous network}
\acrodef{DL}[DL]{deep learning}
\acrodef{DRL}[DRL]{deep reinforcement learning}
\acrodef{EIRP}[EIRP]{effective isotropic radiated power}
\acrodef{FDTD}[FDTD]{Finite-difference time-domain}
\acrodef{FEM}[FEM]{finite element method}
\acrodef{MoM}[MoM]{method of moments}
\acrodef{VNA}[VNA]{vector network analyzer}
\acrodef{CS}[CS]{channel sounder}
\acrodef{CIR}[CIR]{channel impulse response}
\acrodef{CTF}[CTF]{channel transfer function}
\acrodef{DPM}[DPM]{Dirichlet process mixture}
\acrodef{TOA}[TOA]{time of arrival}
\acrodef{GMM}[GMM]{Gaussian mixture model}
\acrodef{OOK}[OOK]{on-off keying}
\acrodef{MLE}[MLE]{maximum likelihood estimation}
\acrodef{LOS}[LoS]{line-of-sight}
\acrodef{NLOS}[NLoS]{non-line-of-sight}
\acrodef{SG}[SG]{signal generator}
\acrodef{SEP}[SEP]{Sun-Earth-probe}
\acrodef{FDSOI}[FDSOI]{fully depleted silicon on insulator}
\acrodef{OpEx}[OpEx]{operational expenditures}
\acrodef{TCO}[TCO]{total cost of ownership}
\acrodef{CapEx}[CapEx]{capital expenditures}
\acrodef{MAC}[MAC]{medium access control}
\acrodef{GEO}[GEO]{geostationary orbit}
\acrodef{SWaP}[SWaP]{size, weight, and power}
\acrodef{NOMA}[NOMA]{Non-orthogonal multiple access}
\ifCLASSINFOpdf
\else
\fi

\begin{document}

\title{AI-Driven Low-Altitude Economy: Spectrum, Mobility, and Validation}

\author{K{\"{u}}r{\c{s}}at~Tekb{\i}y{\i}k,~\IEEEmembership{Member,~IEEE,} 
Amir~Hossein~Fahim~Raouf,~\IEEEmembership{Graduate~Student~Member,~IEEE,} 
\.{I}smail~G{\"{u}}ven{\c{c}},~\IEEEmembership{Fellow,~IEEE,} 
Mingzhe~Chen, G{\"{u}}ne{\c{s}}~Karabulut~Kurt, and
Antoine~Lesage-Landry,~\IEEEmembership{Senior~Members,~IEEE}

\thanks{K. Tekb{\i}y{\i}k, G. Karabulut Kurt, and  A.~Lesage-Landry are with the Department of Electrical Engineering, Polytechnique Montr\'eal \& Poly-Grames Research Centre, Montr\'eal, Canada, (e-mails:\{kursat.tekbiyik, gunes.kurt, antoine.lesage-landry\}@polymtl.ca).}

\thanks{A.~Raouf and \.{I}.~G{\"{u}}ven{\c{c}} are with the Department of Electrical and Computer Engineering, North Carolina State University, Raleigh, NC, 27606, USA (e-mails: \{afahimr, iguvenc\}@ncsu.edu).}

\thanks{M. Chen is with the Department of Electrical and Computer Engineering and Frost Institute for Data Science and Computing, University of Miami, Coral Gables, FL, 33146, USA (email: mingzhe.chen@miami.edu).}

\thanks{This work is supported by the NSERC award ALLRP~579869-22 in Canada and the NSF awards CNS-2332834 and CNS-2332835 in the United States.}
}

\IEEEoverridecommandlockouts 

\maketitle

\begin{abstract}

The Low Altitude Economy (LAE) network, with its transformative capabilities, is a candidate to become one of the major technological developments of the next decade for air mobility. However,  the expected unprecedented density, mobility, and heterogeneity pose challenges and require new approaches, as it renders traditional rule-based approaches inadequate. To address these challenges, this study introduces artificial intelligence (AI)-based approaches and validation frameworks for transitioning AI-enabled technologies from simulation-based studies to practical and deployable systems. This study discusses essential enablers for intelligent LAE networks. First, AI-based spectrum sensing and coexistence utilizing the distributed nature of LAE nodes is introduced.  Then, joint resource allocation and trajectory optimization driven by reinforcement learning is discussed. Bridging the gap between simulation and deployment through experimental platforms such as Aerial Experiments and Research Platform for Advanced Wireless~(AERPAW), which are critical for validating models under realistic and non-stationary airspace conditions, is also addressed. The study concludes by highlighting open issues and outlining a forward-looking roadmap for the development of efficient, interoperable, and scalable AI-driven LAE ecosystems.

\end{abstract}


\IEEEpeerreviewmaketitle
\acresetall

\section{Introduction}\label{sec:intro}

\begin{figure*}[!t]
    \centering
    \includegraphics[width=0.95\linewidth, page=2]{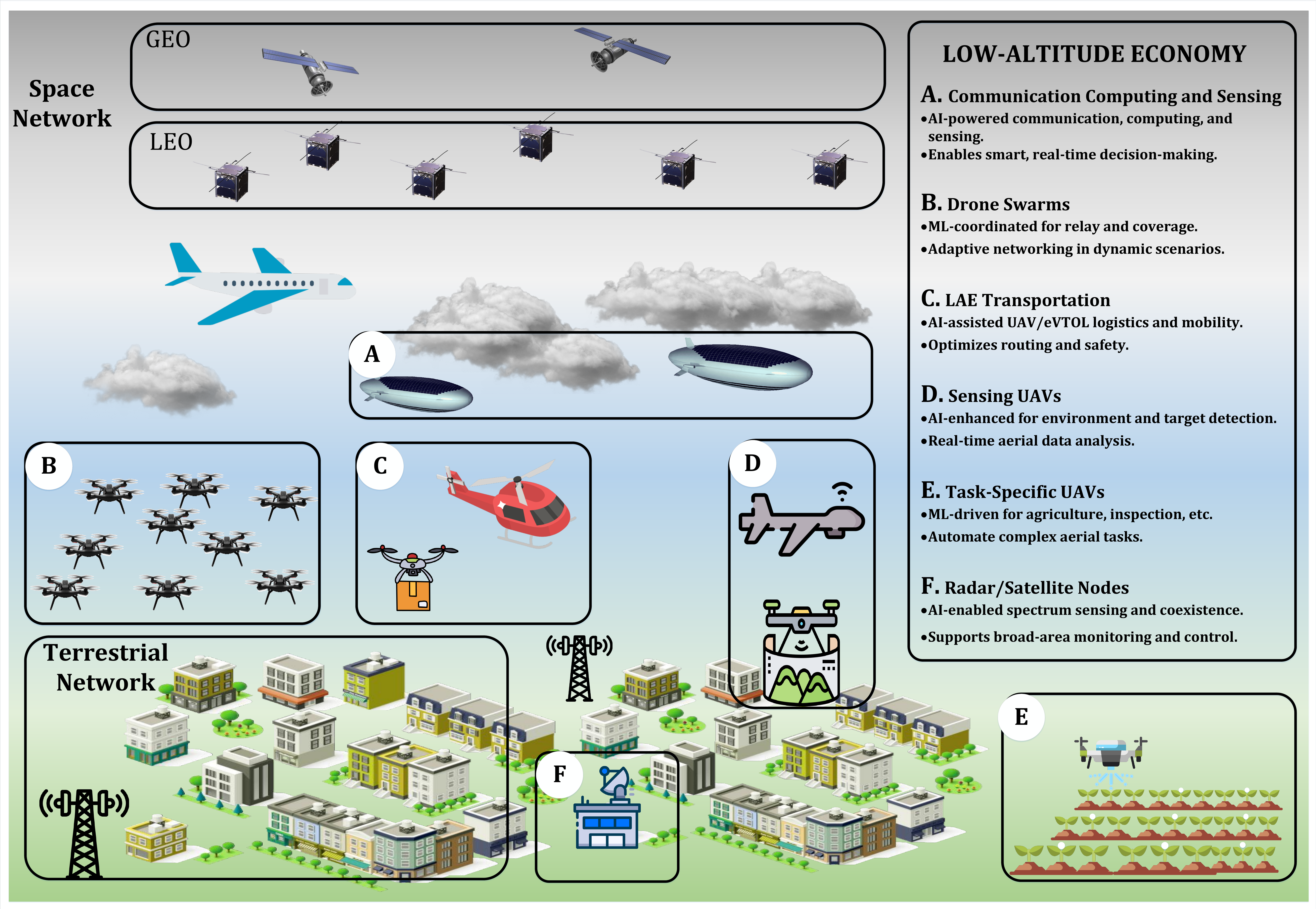}
    \caption{Illustration of an AI-driven Low-Altitude Economy ecosystem integrating intelligent communication, sensing, and computing capabilities. Machine learning enhances coordination, autonomy, and efficiency across diverse aerial operations, supported by seamless air-ground collaboration and spectrum-aware infrastructure.}
    \label{fig:general_view}
\end{figure*}

The Low Altitude Economy (LAE) network is poised to become one of the defining technological trends of the next decade. Encompassing the use of the airspace below $3000$~metres for economic, social, and operational activities, LAE covers various applications: urban air mobility (e.g., air taxis, emergency medical deliveries), precision agriculture, environmental sensing, surveillance, and logistics, as illustrated ixn~\FGR{fig:general_view}. Supported by unmanned aerial vehicles~(UAVs), electric vertical takeoff and landing~(eVTOL) aircraft, high-altitude balloons, and drone swarms, LAE holds enormous promise to revolutionize transportation systems, reduce urban congestion, improve service delivery, and create new economic opportunities. LAE’s importance can be extended beyond pure innovation by addressing critical global challenges such as climate change mitigation, urban sustainability, and disaster resilience~\cite{huang2024potential}. Therefore, LAE can offer an unprecedented opportunity to transform how cities operate and how services are delivered in both urban and remote areas because it enables faster, cleaner, and more flexible aerial operations.

\subsection{Motivations}
Despite the promise of LAE, it presents several challenges, particularly in wireless communications. Unlike traditional terrestrial or satellite networks, LAE networks must contend with unprecedented levels of density, mobility, and heterogeneity. Moreover, the airspace is shared by hundreds or thousands of diverse aircraft, often operating under strict latency, reliability, and energy constraints. These factors collectively introduce significant complexity in managing the wireless spectrum, mitigating interference, ensuring safety, maintaining stable connectivity, and coordinating operations.

Integration of these networks with terrestrial systems adds further complexity. To realize the full potential of LAE, seamless coexistence between airborne, terrestrial, and non-terrestrial networks should be provided. Thus, this requires new strategies for dynamic spectrum access, adaptive channel modelling, robust interference management, and collaborative multi-layer network control~\cite{10955337}. Due to the complex system architecture, needs, and stringent requirements, traditional rule-based approaches can be inadequate for addressing these challenges. Instead, deep learning~(DL)-based approaches can play a significant role in meeting the requirements for safe, efficient, and adaptive LAE networks. Combining dynamic spectrum management, edge computing, real-time spectrum sensing, and multi-agent coordination, it aims to support an ecosystem where diverse platforms can share airspace safely and efficiently. To operate such an ecosystem, coordination across layers is essential, particularly in environments experiencing spectrum congestion and high mobility.

DL offers powerful tools such as distributed and generative learning strategies to enable the harmonious coexistence of air and ground networks. With federated learning~(FL), UAVs can collaboratively sense spectrum usage without central coordination~\cite{tekbiyik2024federated}. Reinforcement learning enables aircraft to dynamically adjust their trajectories and resource allocations in response to environmental changes. Generative models such as conditional generative adversarial networks~(CGANs) help predict radio conditions and optimize channel utilization~\cite{9830137}. The orchestration of these DL techniques promises to make LAE networks more adaptable, resilient, and scalable.

\subsection{Research Focus}
Despite the promising results that DL techniques have shown in theoretical analysis, simulations, or controlled artificial intelligence~(AI) environments using synthetic data, numerous open questions remain before these methods can be effectively implemented in real-world LAE operations. Practical deployments often face challenges such as dynamic spectrum environments, mobility constraints, limited on-board computation, and communication delays. Therefore, this study is dedicated to bridging this gap by examining both the most promising DL-based approaches and the persistent challenges that hinder their practical implementation. In particular, we emphasize the importance of realistic experimentation and validation through advanced testbeds and digital twin platforms such as Aerial Experiments and Research Platform for Advanced Wireless~(AERPAW), which represent a critical step towards operationalizing intelligent, resilient, and scalable LAE networks. \FGR{fig:connections} shows the framework illustrating key challenges in LAE networks, AI-enabled solutions, and their experimental validation. The challenges towards practical LAE systems are introduced in~\SEC{sec:challenges}. Spectrum scarcity and dynamic interference management motivate distributed spectrum sensing addressed in~\SEC{sec:spectrum_management}. Also, high mobility and 3D heterogeneous networks require adaptive integration of communications, computing, and sensing (C2S) together with trajectory-aware optimization discussed in~\SEC{sec:resource_allocation}. Finally, due to the lack of realistic datasets and standardized evaluation environments, empirical validation is a must, which is the core of~\SEC{sec:testbed} through testbed-driven experimentation and data collection. This challenge–enabler–validation loop proposes a systematic roadmap for transitioning AI-enabled LAE technologies from simulation-based studies to practical and deployable systems.

\begin{figure*}[!t]
    \centering
    \includegraphics[width=0.95\linewidth, page=6]{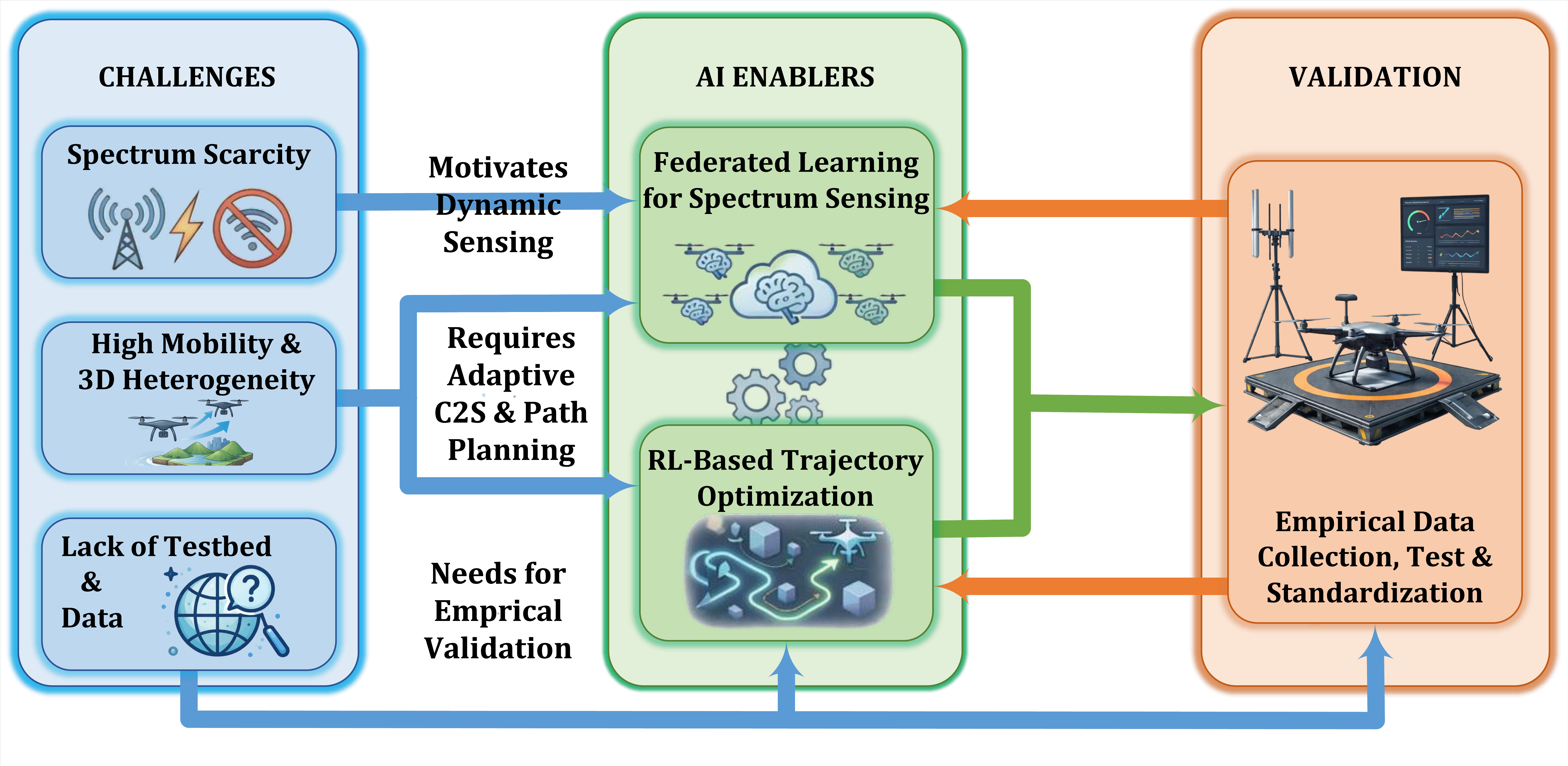}
    \caption{Framework illustrating the interaction between LAE challenges, AI enablers, and experimental validation toward deployable intelligent aerial networks.}
    \label{fig:connections}
\end{figure*}

\section{Challenges in Low-Altitude Wireless Networks}\label{sec:challenges}

The growth in LAE systems introduces significant challenges across multiple technical aspects. Although numerous challenges exist, such as air traffic management, energy limitations, security risks, privacy concerns, and regulatory compliance~\cite{10955337}, this study focuses specifically on spectrum coexistence,  integration of C2S, and standardization and testbed, which are addressed in subsections below. 

\subsection{ML in Spectrum Utilization and Coexistence}\label{sec:spectrum_utilization}

Due to the highly dynamic and dense nature of low-altitude airspace, optimizing spectrum utilization presents major challenges. For example, high mobility among UAVs and eVTOLs leads to Doppler shifts, frequent handoffs, and time-varying multipath fading, significantly degrading link stability. Another issue is that interference becomes particularly inevitable in urban environments where aerial vehicles must coexist with terrestrial base stations, Wi-Fi networks, legacy aviation systems, and radars.  Therefore, how to dynamically sense and allocate spectrum resources, predict interference patterns in real time, and balance among competing air and ground users are the main questions remaining open, yet. 

ML offers opportunities to optimize spectrum utilization and interference management in LAE networks; however, its practical application requires solid solutions for several challenges. One of the most important issues is the non-stationary and spatially correlated nature of the environment. Unlike terrestrial networks, LAE consists of highly mobile platforms such as UAVs and eVTOLs operating at different altitudes and media. As a result, it leads to changing signal propagation conditions, including frequent transitions between line-of-sight~(LoS) and non-line-of-sight~(NLoS) links, and fluctuations in interference levels. ML models trained on localized or static datasets have a risk of failing to generalize across the entire 3D airspace. This might reduce the reliability of spectrum estimation and classification.

Another important challenge is the scarcity of labelled and reliable data required to train robust spectrum sensing models. In LAE environments, acquiring ground truth information about spectrum occupancy or interference is not only costly but also sometimes impractical. For example, military radar systems or special emergency broadcasts may be difficult to detect or label accurately. 

Moreover, cross-layer dynamics create additional complexity. Spectrum utilization in LAE is inherently a cross-layer problem because physical layer measurements, such as signal-to-noise ratio~(SNR) and Doppler shift, interact with medium access control (MAC) and network layers.  ML models trained in isolation without timely feedback from other layers can become out of sync with the operational context, which is known as model drift~\cite{10163748}. In practice, synchronization between layers is difficult to achieve because of limited cross-platform communication and the asynchronous nature of distributed LAE nodes. This can reduce spectrum reuse efficiency or harm coordination in dynamic spectrum access.

Finally, the lack of standardized simulation environments and datasets tailored to LAE-specific conditions significantly delays progress. Most existing ML models for spectrum estimation or interference mitigation are evaluated under simplified or terrestrial scenarios that fail to capture the 3D and time-varying characteristics of low-altitude networks. In the absence of common metrics, reproducibility is limited, and the generalizability of findings remains uncertain. Therefore, regulation and standardization efforts are essential to establish a reliable ML-based spectrum management infrastructure. Moreover, the blackbox decision-making processes of many ML models limit their reliability and regulatory acceptance, particularly in spectrum-sensitive domains such as aviation, where explainability and robustness are prerequisites for deployment. As a result, the integration of ML into LAE spectrum systems is likely to remain an open challenge until these domain-specific constraints are effectively addressed. As we discuss later, digital twin systems that mimic the real environment and experiment platforms are the key enablers for developing and testing these kinds of methods. They accelerate and make the progress toward practical ML model deployments more efficient.

\subsection{Integration of C2S}\label{sec:c2s}

While effective spectrum utilization and interference management are important to connectivity in LAE networks, they show only one dimension of a broader challenge. The integration of C2S in LAE networks presents a set of technical challenges, such as the need to jointly optimize diverse functions under high mobility and resource constraints. In LAE, UAVs are expected to simultaneously operate as mobile edge computing platforms, base stations, relays, and sensors~\cite{10536071}. This results in competing demands on limited onboard resources such as energy, computational capacity, and, of course, link capacity. Coordinating these functions in real time is particularly challenging owing to changing spectrum conditions and low-latency application demands such as delivery and surveillance.

A critical challenge is the scheduling of C2S tasks across a swarm of heterogeneous aerial nodes. For instance, high-resolution sensing workloads can saturate data pipelines so that communication bottlenecks can delay time-sensitive decisions such as path changes or mission replanning. The lack of unified control systems might cause suboptimal performance and poor resource utilization. Heterogeneity in UAV platforms, e.g., varying in computational capacity, sensor types, and antenna systems, etc., makes synchronized data fusion and task assignment further challenging~\cite{10955337}.

Finally, a key challenge is the lack of standardized architectures that can support scalable, fault-tolerant C2S integration. Current implementations tend to stack functions vertically rather than holistically design coordinated workflows with feedback across layers. Bridging this gap requires not only algorithmic developments but also robust system designs that consider the interactions among the physical layer, the network, and sensors~\cite{wang2019survey}. Along with these, proper test environments accurately simulating real-world scenarios, including the interactions between the physical layer and upper layers as well as hardware characteristics, are required.

\subsection{Standardization and Testbed}\label{sec:standardization}

Deployment of LAE networks requires a complex standardization landscape that imposes significant challenges. Therefore, regulators such as Ofcom aim to establish trust and connectivity by shifting generic license exemptions to specific UAV radio licenses, thereby creating interference-protected airspaces~\cite{ofcom_uas_2022}. Also, the Federal Aviation Administration (FAA) introduces urban air mobility corridors to manage air traffic with AI agents learning from dynamic and community-based rules rather than rigid clearances.  For the full integration and autonomy,  the International Civil Aviation Organization (ICAO) proposes a new framework that enables AI systems handle frequent decisions currently limited by human reaction times to harmonize altitude and separation standards globally~\cite{icao_utm_framework_2023}. However, these visions require experimental validation, yet the physical infrastructure to bridge the gap between these conceptual standards and deployment remains underdeveloped.

Beyond challenges in spectrum management and integrated C2S functionality, the improvement of LAE networks has been significantly limited by the lack of standards and real-world testbeds. Existing standards developed for terrestrial 5G, satellite communications, and manned aviation cannot meet the operational and regulatory demands specific to low-altitude platforms. 

In parallel, the limited publicly available testbeds undermine the validation and benchmarking of proposed systems and algorithms. Although simulation remains useful for preliminary analysis, it fails to capture some physical layer nuances of LAE environments, such as altitude-dependent multipath fading, spatiotemporal spectrum utilization, and platform-dependent distortions. As detailed later, to address this, the AERPAW testbed has emerged as a critical resource enabling controlled experiments with programmable UAVs and software-defined radios~(SDRs) in mixed urban-rural landscapes~\cite{marojevic2020advanced}. AERPAW supports the evaluation of AI-driven spectrum access and cross-layer optimization under realistic constraints, bridging theory and practice, and hosts over two dozen datasets based on real-world UAV network environments.

However, several challenges still exist. Reproducing dense urban conditions with high fidelity, simulating interference or jamming scenarios, and scaling the testbed infrastructure to accommodate a variety of aerial platforms, such as eVTOLs or fixed-wing drones, are non-trivial tasks. Moreover, the simulation-to-reality gap persists; ML models trained on digital twins often underperform in real-world deployments due to unpredictable environmental conditions~\cite{10945722}. To accelerate LAE adoption, future efforts should prioritize interoperability across testbeds and the development of standardized evaluation metrics. Additionally, collaboration among researchers is essential for sharing test data, harmonizing spectrum policies, and establishing certification protocols.

\section{AI-Driven Spectrum Management and Interference Mitigation}\label{sec:spectrum_management}

\begin{figure*}[!t]
    \centering
    \subfigure[Transmit power]{
        \label{fig:acc_vs_tx_power}
        \includegraphics[width=0.4\linewidth]{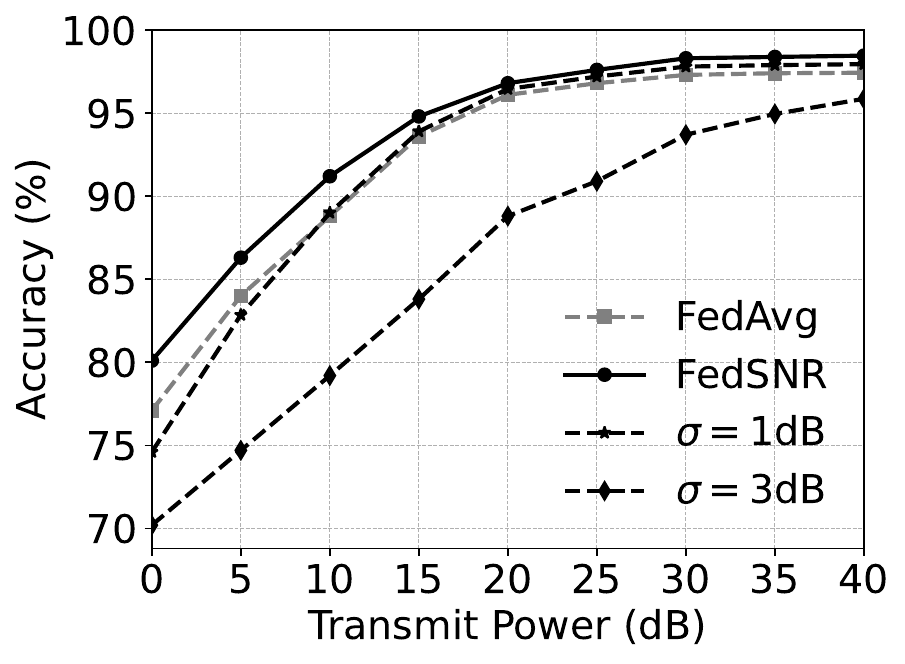}}
    \quad
    \subfigure[Dataset size per UAV]{
        \label{fig:acc_vs_batch}
        \includegraphics[width=0.4\linewidth]{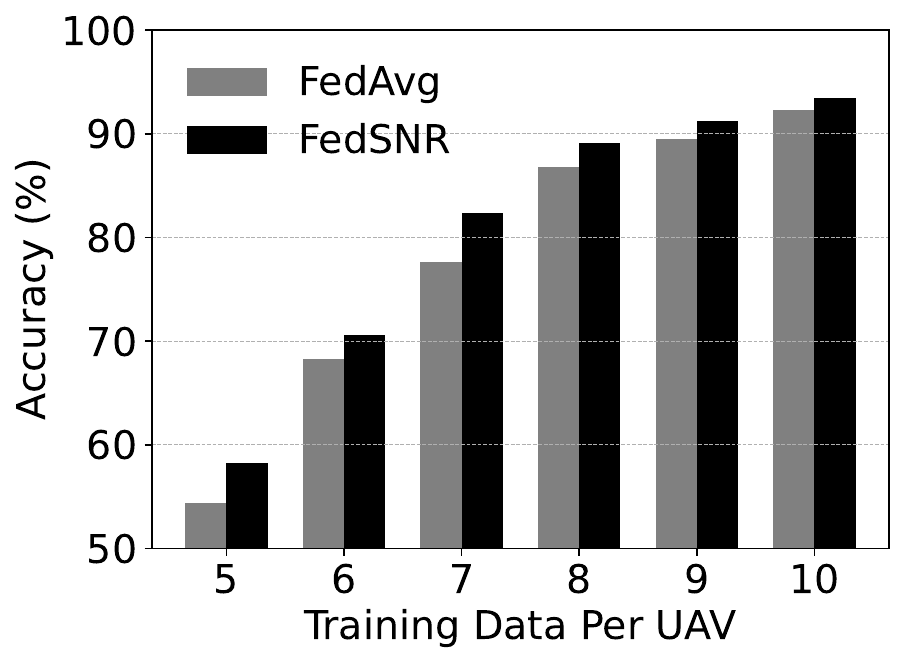}}
    \caption{FL-based spectrum sensing performance results.}
    \label{fig:fl_ss}
\end{figure*}

Spectrum management and interference detection constitute key enablers for robust LAE network operations. Because aerial systems become denser and more dynamic, the ability to intelligently sense, allocate, and manage spectrum resources is essential. AI techniques, particularly ML, have emerged as powerful tools to address these challenges, offering scalable, adaptive, and data-driven approaches. Accordingly, this section discusses ML-based approaches for spectrum sensing and interference detection.  

\subsection{AI-Enabled Dynamic Spectrum Sensing for Low-Altitude Networks}\label{sec:spectrum_sensing}

Dynamic spectrum sensing is a fundamental component for enabling reliable and efficient communication in highly congested LAE, where aerial and terrestrial users share limited frequency resources. In~\cite{tekbiyik2024federated}, we proposed an FL method to address the challenges of distributed UAV networks, where centralized solutions are impractical because of power constraints, privacy concerns, and high communication overhead. The proposed system utilizes a lightweight convolutional neural network (CNN) architecture, which is specifically designed to operate on resource-constrained UAV platforms. By processing the real and imaginary components of received signals using one-dimensional convolutions, the system captures temporal signal patterns while trying to maintain computational efficiency. A key contribution of the study is the introduction of the FedSNR aggregation method, extending the conventional federated averaging (FedAvg) approach by incorporating the SNR observed at each UAV into the aggregation process. This approach allows the global model to give greater weight to updates from UAVs experiencing higher-quality signals, thereby enhancing adaptability, which is particularly important in heterogeneous and dynamic environments.

The experimental setup in~\cite{tekbiyik2024federated} involves UAV networks operating in a realistic 3D space, with radar systems generating various waveform types, including continuous wave, chirp, and pulse signals. The CNN is carefully designed with batch normalization and dropout layers to improve training stability and generalization. A global average pooling layer is employed to reduce model complexity and memory requirements, enabling deployment on resource-constrained UAVs on edge devices. Numerical simulations demonstrate the proposed system, with FedSNR achieving up to a $15\%$ improvement in sensing accuracy over centralized models~\cite{tekbiyik2024federated}. Notably, as the number of UAVs increased, the system’s sensing accuracy improves. It shows the benefit of incorporating multi-channel spatial diversity in the learning process. Although the proposed approach outperforms FedAvg, particularly in low SNR regimes as indicated in~\FGR{fig:acc_vs_tx_power}, we further test the framework by adding error to the SNR used in FedSNR aggregation because gathering the actual SNR might be challenging in practice, and the estimated SNR might include error, of course. To simulate this kind of error-prone SNR estimation, we feed the SNR values into the aggregation process by adding Gaussian noise, $\mathcal{N}(0, \sigma^2)$, to the real SNR. As seen in~\FGR{fig:acc_vs_tx_power}, erroneous SNR estimations hinder FedSNR's learning, especially at low transmit power. Thus, these results point to a gap in the use of the FedSNR method in practice and require a good SNR estimator with an error around $\sigma = 1$~dB. Then, \FGR{fig:acc_vs_batch} shows a critical issue in distributed learning when data scarcity causes a weight divergence problem because environmental conditions and mobility patterns vary widely between nodes, yielding generally a non-independent and identically distributed (non-IID) dataset. This situation highlights why the real environment and digital twin validation are sine qua non for distributed systems, because the simulations might overlook the impact of SNR estimation error,  mobility, and non-IID data distributions on the system performance. However, these practical challenges lead to model divergence and degrade performance under data scarcity. As a result, this gap remains open, and addressing it requires the use of experimental platforms like AERPAW, as we discuss in the following sections. These should be carefully investigated in future studies.

In summary, the proposed FL framework for UAV-based spectrum sensing offers a privacy-preserving and communication-efficient solution. By joint utilization of a distributed lightweight CNN model with FedSNR aggregation strategy, the system addresses accurate sensing and resource constraints in LAE networks. However, the issues discussed above indicate crucial open points towards practical applications. These outcomes provide an important insight for future research and practical implementations in real-world aerial communication systems, where spectrum-sharing requirements demand adaptive and efficient learning solutions.

\subsection{ML Techniques for Interference Detection and Mitigation}\label{sec:interference_detection}

Beyond spectrum sensing, ML plays an essential role in active interference management. Techniques such as reinforcement learning (RL) can allow UAVs to dynamically adjust their transmission power, channel selection, and flight trajectories to minimize interference and maximize throughput. Also, generative models such as 3D-CGANs~\cite{9830137} can further enhance interference mitigation by predicting spatiotemporal channel occupancy maps, enabling UAVs to proactively avoid congested channels. This predictive capability is valuable in dense urban environments because the radio frequency medium can change rapidly due to UAV mobility and reflections from buildings. \looseness = -1

Future research directions can include the development of hybrid models that integrate FL with RL to create systems that combine local adaptability with global coordination. Multi-agent RL techniques enable UAV swarms to cooperatively learn interference patterns~\cite{wu2019fundamental}. In real applications, it is important to balance computing power, delay, and energy use so that ML methods do not exceed the limited resources of aerial vehicles unlike the current literature trying to optimize only one of them.

\section{AI-Optimized Resource Allocation and Trajectory Planning}\label{sec:resource_allocation}

Following intelligent spectrum utilization and interference mitigation methods, efficient resource allocation and trajectory planning are a must for stable LAE. The challenge is that these tasks demand real-time decisions over multidimensional resources while having uncertain and dynamic environments. Unlike terrestrial systems, because LAE networks are characterized by high mobility, strict energy and delay constraints, a holistic design approach jointly considering UAV trajectory control and communication resource scheduling is demanded.  Hence, this section discusses the possible solutions regarding joint resource allocation and trajectory planning.

\subsection{Reinforcement Learning–based Resource Management in Dynamic Airspace}\label{sec:rl_resource_managenment}

Optimizing UAV trajectories is essential for reducing electromagnetic interference and improving power efficiency. Conventional preplanned trajectories often fail in highly dynamic environments where network conditions and interference frequently evolve.  Also, optimizing only communication performance without considering power consumption and mobility leads to suboptimal results. There exist essential trade-offs among throughput, delay, and power consumption that should be jointly addressed to obtain effective LAE systems. For example, improving communication throughput by flying airborne closer to users requires additional propulsion energy, while reducing delay by speeding up may also rapidly consume limited onboard energy~\cite{wu2019fundamental}. 

Compared to terrestrial base stations, aerial nodes can dynamically reposition to maintain LoS links, adapt to ground user movement, and exploit location-dependent channel conditions. But every repositioning decision also impacts delay and power consumption. For example, hovering closer to users can improve SNR but consumes excessive power for rotary-wing UAVs. Similarly, fixed-wing UAVs that are efficient for wide-area scanning but cannot hover require periodic trajectory planning.

\looseness = -1
Deep reinforcement learning and trajectory-aware resource optimization might be well-tailored to address these multidimensional trade-offs in practical applications. The RL life cycle for LAE resource optimization starts with the UAV observing its environment, including its position, channel state, and weather. Based on a learned policy, it selects an action like adjusting its trajectory or adjusting transmit power. After executing the action, it receives a reward for performance and stability. This experience updates the policy through iterative learning, often using techniques like experience replay and target networks. Over time, this loop of sensing, acting, and learning enables UAVs to adapt to dynamic airspace conditions efficiently.

For example, in~\cite{10891537}, it is shown that the digital twin network (DNT) combining gated recurrent units (GRUs) with value decomposition networks (VDN) enables distributed agents like UAVs and base stations to collaboratively optimize spectrum and computing resources. This approach outperforms independent Q-learning baselines, showing the power of cooperative learning. This is especially useful in LAE environments where manual design of reward functions is difficult due to complex, dynamic objectives.

\subsection{Optimization of UAV Trajectories to Minimize Interference}\label{sec:trajectory_optimization} 

In multi-UAV networks or dense deployments, the challenge is further amplified due to intra-system or inter-system interference. Optimizing spectrum reuse and user association across systems requires coordinated trajectory planning and power control.  RL algorithms, such as deep Q-networks, deep deterministic policy gradients, and multi-agent reinforcement learning, allow autonomous adjustment of the flight paths based on real-time feedback. These algorithms can optimize routes to avoid high-interference zones, align with directional beams, and maintain stable links even in dense urban environments.

Moreover, trajectory optimization can improve fairness and quality of service by balancing coverage across users and reducing interference~\cite{gao2024federated}. However, many works assume idealized channel and interference models or simple network topologies which may not fully capture dynamic mobility and heterogeneous nodes. Another important point missed by these works is nonideal flight dynamics consisting of no‐fly zones, regulatory and safety constraints, 3D mobility, weather, changes in ground network loads. Thus, bridging to the real world is still a challenge.

\section{Experimental Validation and Testbed Deployment}\label{sec:testbed}

Besides the AI-driven frameworks and optimization strategies discussed in the previous sections, real-world validation is essential. Experimental validation not only demonstrates the feasibility of theoretical models but also bridges the gap between simulation and practical deployment and provides critical insights for practical LAE networks. In this section, we introduce AERPAW along with performance metrics regarding the validation of LAE systems and some practical insights learned from the testbed for field-deployments.

\begin{figure*}[!t]
    \centering
    \includegraphics[width=0.95\linewidth]{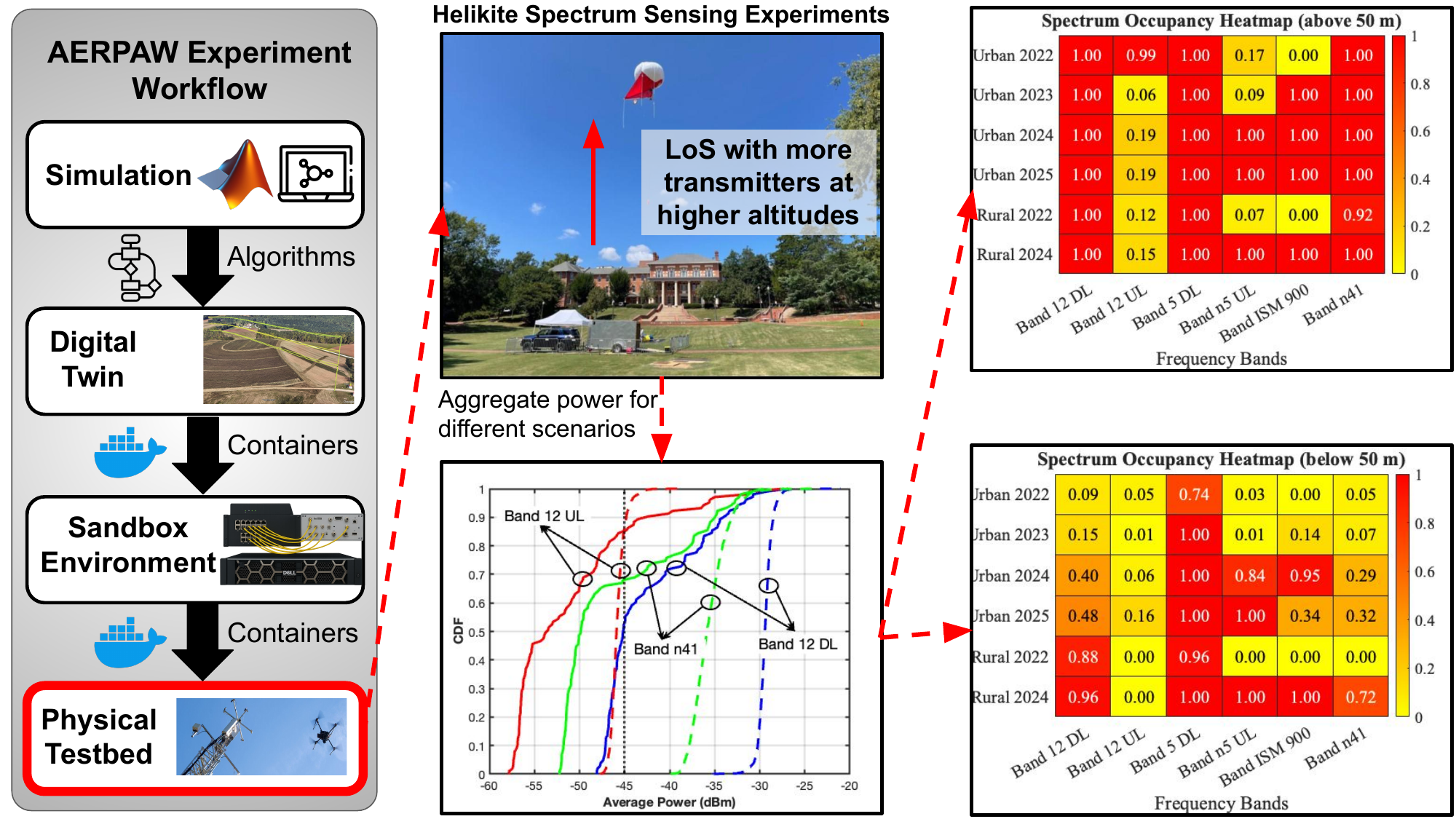}
    \caption{The AERPAW platform supports three operational modes: physical testbed, sandbox, and digital twin environments. In the CDF plot, solid and dashed curves correspond to measurements taken below and above 50~m altitude, respectively. The vertical black dotted line represents the threshold value of $-45$~dBm, which is used to determine the spectrum occupancy across frequency bands.}
    \label{fig:aerpaw}
\end{figure*}

\subsection{AERPAW: An Experimental Platform for LAE Research}

AERPAW, located in Raleigh, NC, serves as a flagship testbed within the NSF Platforms for Advanced Wireless Research (PAWR) program and is uniquely tailored for LAE experimentation. AERPAW, consisting of programmable UAVs and ground nodes equipped with SDRs, offers a highly flexible platform for controlled wireless experiments in heterogeneous environments~\cite{marojevic2020advanced}. By spanning both urban and rural areas e.g., Lake Wheeler Field Labs and the NC State Centennial Campus, AERPAW enables reproducible and scalable experiments for spectrum management strategies and trajectory optimization techniques discussed above.

As AERPAW combines live outdoor experimentation and digital twin–based emulation, it provides a powerful framework for researchers to design, simulate, and validate aerial wireless communication scenarios before field deployment. This capability is critical to validate the methods and architectures towards regulation, standardization, and practical usage because it allows for iterative refinement of LAE systems.

As illustrated in~\FGR{fig:aerpaw}, other than simulations that are often used in evaluating LAE networks, the AERPAW platform supports three experimental modes. The digital twin mode enables users to develop and test code virtually on cloud-based nodes, offering flexibility but limited physical realism. The sandbox mode employs real radio hardware in controlled indoor environments, allowing scheduled access to live radios without UAV mobility. The testbed mode executes experiments on outdoor nodes in the field, where AERPAW personnel conduct the tests on behalf of users, without live interaction.

\subsection{Deployment Insights and Testbed Lessons}

\looseness = -1
The practical experience with AERPAW yields several important lessons for LAE development. First, trajectory-aware communication strategies are indispensable under the changing propagation conditions associated with UAV mobility, altitude, and orientation. Second, bridging the simulation-to-reality gap remains a nontrivial challenge. Models trained in digital twin environments often require calibration with real-world measurements to achieve reliable operation. Finally, the modular design of AERPAW's portable radio nodes and its abstraction layers enables flexible, future-proof architectures capable of accommodating heterogeneous hardware and regulatory requirements.

\FGR{fig:aerpaw} presents representative results illustrating altitude-dependent spectrum characteristics in rural and urban environments for LAE networks. A helium-assisted kite~(helikite) was employed to collect wideband spectrum measurements between 85~MHz and 6~GHz at multiple altitudes, locations, and measurement periods. The empirical cumulative distribution function~(CDF) of the average received power from the Urban 2025 measurement campaign is presented for LTE downlink Band~12, LTE uplink Band~12, and 5G Band~n41, with results grouped by UAV altitudes below and above 50~m. The CDFs reveal that at lower altitudes, received signal power is generally weaker, whereas higher altitudes yield noticeably stronger levels due to improved LoS conditions, with most samples exceeding the $-45$~dBm threshold.

The same figure includes heatmaps depicting spectrum occupancy across several frequency bands of relevance to LAE networks, comparing urban and rural environments below and above 50~m altitude. For each band and altitude group, occupancy is determined from the mean received power across the corresponding band, which is considered occupied when this mean exceeds $-45$~dBm. At lower altitudes, occupancy patterns vary considerably; for instance, LTE downlink Band~12 shows limited activity in Urban~2022 but significant utilization in Rural~2024. At higher altitudes, spectrum occupancy becomes denser and more uniform across most bands and years. These results confirm that both signal availability and interference conditions are strongly influenced by altitude, emphasizing the importance of altitude-aware spectrum characterization for reliable and efficient LAE network deployment.

The unique characteristics and statistics of the real-world data, such as for altitude-dependent spectrum occupancy, underscore the importance of AI-driven solutions presented in this study. Such data-driven AI solutions can be used to improve spectrum sensing, interference mitigation, resource allocation, and trajectory optimization for LAE networks. They also highlight the need for integrated C2S frameworks and standardized testbed practices. Specifically, these results bridge the gap between the theoretical C2S challenges introduced in~Section~II-B and practical deployment. For instance, the altitude-dependent spectrum heterogeneity shown in the heatmaps implies that the radio environment is too dynamic, and therefore, static rule-based models could struggle. A collaborative C2S framework must integrate such real-world sensing capabilities to allocate resources. These experiments validate the need for a holistic C2S design for LAE networks because the results illustrate that physical sensing constrains AI models.

Moreover, these measurements provide real-world data required to train and validate the AI models discussed in the previous sections. For example, the altitude-dependent path loss and interference data collected from the testbed are used to train Physics-Inspired Kolmogorov-Arnold Networks (PIKAN)~\cite{tekbiyik2025pikan}. PIKAN leverages this real-world dataset to learn symbolic, interpretable expressions of the channel that also employ physical propagation laws. This demonstrates that the AERPAW setups act as an enabler for  AI model training and validation to capture complex non-stationary airspace behaviors such as elevation and altitude-dependent fading, while theoretical simulations often oversimplify these impacts.

AERPAW plays a critical role in data generation, model validation, and standardization efforts for LAE networks. The testbed can serve as a validation platform for emerging protocols related to spectrum coexistence, trajectory planning, and resource optimization. By supporting standardized interfaces and benchmarks, AERPAW can help ensure that AI-enhanced solutions are not only technically feasible but also interoperable and aligned with evolving global standards.

\section{Open Issues}\label{sec:open_issues}

Although we mention some open issues specific to research areas given above, this section aims to discuss open issues from a broader perspective, without narrowing to specific tasks such as spectrum sensing. We aim to provide holistic future directions toward AI-driven LAE.

As LAE networks integrate into critical infrastructure, the security of their AI-driven components needs to be a crucial concern. As we explained, distributed frameworks like FL can reduce data leakage, but model poisoning and adversarial attacks might manipulate the global model. In spectrum sensing or trajectory planning, a malicious edge node could change training weights to mask interference or induce collisions. Consequently, deploying AI in LAE requires sophisticated aggregation protocols and anomaly detection mechanisms so that these systems can improve safety and maintain operational integrity. However, the primary concern is the integration of explainable AI and trustworthiness techniques into decision-making frameworks. As AI models impact mission-critical operations, it becomes important to ensure transparency and interpretability in their decisions, especially under uncertain or adversarial conditions. To address this, recent frameworks such as PIKAN~\cite{tekbiyik2025pikan} is proposed to bridge the gap between DL accuracy and physical interpretability. Drawing inspiration from physical principles such as path loss exponents and two-ray models, PIKAN learns channel behaviour and provides symbolic expressions that explain the underlying logic of the predictions. Such physics-inspired and explainable approaches are required for certification and regulatory approval because aviation authorities audit and trust the decisions made by autonomous aerial systems based on those.

Another issue is the need for developing unified metrics for performance evaluation of AI models in LAE scenarios. Existing datasets and performance metrics can fail to capture the unique dynamics of low-altitude airspace, such as 3D mobility, heterogeneous systems, and dynamic spectrum availability. Realistic simulation environments and standardized evaluation protocols can produce meaningful progress and fair comparisons between different approaches.

Besides, establishing interoperable standards between air, space, and terrestrial systems remains a challenge. Cooperation with satellite systems has been studied in the literature, but more efforts are needed for seamless integration that requires well-defined protocols and spectrum policies.

Last, practical concerns such as energy-efficient AI models, robust operation in harsh environments, and real-time adaptation to dynamic conditions require further investigation. Experimental testbeds, FL frameworks, shared datasets, and open-source platforms can help bridge the gap between theoretical advances and real-world deployment. These efforts would also improve community-driven innovation and reduce the risks of commercial adoption of LAE technologies.

\section{Concluding Remarks}\label{sec:conclusion}

The emergence of LAE networks presents an exciting frontier in wireless communications as offering transformative capabilities across several domains such as aerial transportation, environmental monitoring, and logistics. However, realizing these promising capabilities requires overcoming challenges in communication, sensing, and system coordination within highly dynamic and resource-constrained airspace environments.
In this study, we provide a comprehensive exploration of the key wireless communication subsystems required to support the vision of AI-driven LAE networks. Along with detailing the challenges towards full operable LAE networks, we identify and discuss three critical issues as follows: spectrum sensing and coexistence strategies empowered by ML, RL-driven resource allocation and trajectory planning, and the standardization and experimental validation of proposed solutions through real-world testbeds.

ML-based approaches can provide promise in adapting to the non-stationary, heterogeneous conditions of LAE. For instance, FL models, particularly those incorporating SNR-aware aggregation schemes, can support decentralized and privacy-preserving spectrum sensing while maintaining high accuracy in evolving environments. It should be noted that there are several open issues for practical deployment. We also highlight the role of experimental platforms like AERPAW in narrowing the gap between theoretical innovations and practical deployments. Through digital twins, sandbox environments, and live testbeds, they can enable validation of AI-supported LAE solutions under realistic conditions. These efforts are essential not only for technical development but also for ensuring regulatory compliance, interoperability, and robustness across diverse deployment scenarios. 

In summary, AI-driven LAE networks show a convergence of wireless communications, edge intelligence, and autonomous control. By developing ML-based methods, validating them through realistic experiments, and aligning them with standardization efforts, the wireless research community can shape the next generation of safe and intelligent aerial systems.

\balance

\bibliographystyle{IEEEtran}
\bibliography{main}

%\vspace{-0.2cm}

%%%%%%%%%%%%%%%%% BIOGRAPHIES %%%%%%%%%%%%%%%%%%%%%%
\section*{Biographies}
\footnotesize{K{\"{U}}R{\c{Ş}}AT TEKBIYIK [StM'19, M'24] (kursat.tekbiyik@polymtl.ca) received his Ph.D. degree in telecommunications engineering from Istanbul Technical University, Istanbul, T{\"{u}}rkiye. His research interests include machine learning applications in wireless communications, terahertz communications, and non-terrestrial networks.\\

AMIR HOSSEIN FAHIM RAOUF [StM'16] (afahimr@ncsu.edu) is currently pursuing a Ph.D. degree in Electrical Engineering at North Carolina State University, Raleigh, NC, USA. His main research areas are wireless communications, visible light communications, and quantum key distribution. \\

\.{I}SMA\.{I}L G{\"{U}}VEN{\c{C}} [F'20] (iguvenc@ncsu.edu) received his Ph.D. in electrical engineering from the University of South Florida in 2006. He is currently a professor at NC State University, Raleigh, NC.  \\

MINGZHE CHEN [StM'15, M'19, SM'24] (mingzhe.chen@miami.edu) is currently an Assistant Professor with the Department of Electrical and Computer Engineering at University of Miami. His research interests include federated learning, reinforcement learning, virtual reality, unmanned aerial vehicles, and Internet of Things.\\

G{\"{U}}NE{\c{Ş}} KARABULUT KURT [StM'00, M'06, SM'15] (gunes.kurt@polymtl.ca) is Professor
of Electrical Engineering at Polytechnique Montréal, Montréal, QC, Canada. She is a Marie Curie Fellow and has received the Turkish Academy of Sciences Outstanding Young Scientist (TÜBA-GEBIP) Award in 2019. She received her Ph.D. degree in electrical engineering from the University of Ottawa, ON, Canada.\\

ANTOINE LESAGE-LANDRY [StM'16, M'20, SM'24] is an Associate Professor of Electrical Engineering at Polytechnique Montréal, QC, Canada. His research focuses on decision-making in complex engineering systems like electric power systems and wireless communication networks.
}
%%%%%%%%%%%%%%%%% BIOGRAPHIES %%%%%%%%%%%%%%%%%%%%%%

\balance
\end{document}